**[Title Page]**

Article Title

**MPO: An Efficient and Low-cost Peer-to-Peer Overlay for Autonomic Communications**

**Authors**   Jiaqi Liu[a,*], Zhong Ren[b], Deng Li[c]

**Author affiliations**

[a] School of Software, Central South University, 410075 Changsha, China

[b] Graduate School of Natural Science and Technology, Kanazawa University, Japan

[c] School of Information Science and Engineering, Central South University, 410083 Changsha, China

**Correspondence information: Jiaqi Liu, Central South University,**

**jiaqi1006@gmail.com, +86 13723875947**

**(Check the Guide for authors to see the required information on the title page)**

[*] Corresponding author. Tel.: +86 13723875947; E-mail address: jiaqi1006@gmail.com

# MPO: An Efficient and Low-cost Peer-to-Peer Overlay for Autonomic Communications

Jiaqi Liu[a,†], Zhong Ren[b], Deng Li[c]

[a] School of Software, Central South University, 410075 Changsha, China

[b] Graduate School of Natural Science and Technology, Kanazawa University, Japan

[c] School of Information Science and Engineering, Central South University, 410083 Changsha, China

**Abstract**

The term Autonomic Communication (AC) refers to self-managing systems which are capable of supporting self-configuration, self-healing and self-optimization. However, information reflection and collection, lack of centralized control, non-cooperation and so on are just some of the challenges within AC systems. We have considered these problems in theory and practice and reached the following conclusion; in order to build an ideal system for autonomic communication, there are three key problems to be solved: (1) how to maintain system structure stability in front of the dynamic topologies; (2) how information can be located and routed at a low cost without global view of the system; (3) how to make the overlay robust. Motivated by the need for AC, we have designed an efficient and low-cost Peer-to-Peer (P2P) overlay called Maya-Pyramid overlay (MPO) and combined merits of unstructured P2P with those of structured P2P overlays. Differing from the traditional hierarchical P2P (i.e.

---

[†] Corresponding author. Tel.: +86 13723875947; E-mail address: jiaqi1006@gmail.com

tree-like structure) overlay, (1) MPO is composed of levels and layers, which uses small world characteristic to improve efficiency, and the maintenance cost is decreased because update and backup only take place in two neighboring levels or layers instead of recursively perform in higher levels. (2) Unlike normal redundant mechanisms for solving the single fault problem: Tri-Information Center (Tri-IC) mechanism is presented in order to improve robustness by alleviating the load of cluster heads in a hierarchical P2P overlay. (3) A source ranking mechanism is proposed in order to discourage free riding and whitewashing and to encourage frequent information exchanges between peers. (4) Inspired by Pastry's ID structure for a structured DHT algorithm, a 3D unique ID structure is presented in the unstructured P2P overlay. This will guarantee anonymity in routing, and will be, not only more efficient because it applies the DHT-like routing algorithm in the unstructured P2P overlay, but also more adaptive to suit AC. Evaluation proved that MPO is robust, highly efficient and of a low-cost.

MPO: An Efficient and Low-cost Peer-to-Peer Overlay for Autonomic Communications

# 1. Introduction

The term Autonomic Communication (AC) is used for this form of self-managing systems able to support self-configuration, self-healing and self-optimization—the so-called self-* properties. AC is a new paradigm to assist the design of the next generation networks: self-organizing, context-aware pervasive systems and service-oriented computing environments so as to better support highly dynamic and mobile users and virtual organizations. Such data-intensive, unstructured environments with minimal or no centralized control present a challenge for traditional methods of analysis and design. Since many self-* properties (e.g. self-configuration, self-optimization, self-healing, and self- protecting) are achieved by a group of autonomous entities that coordinate in a peer-to-peer (P2P) fashion, thus, it has opened the door to migrating research techniques from P2P systems. P2P's meaning can be better understood with a set of key characteristics similar to AC: decentralized organization, self-organizing nature (i.e. adaptability), resource sharing and aggregation, and fault-tolerance. Moreover, there are a number of critical challenges and problems which currently prevent the great potential of AC from being revealed: such as information reflection and collection, lack of centralized control, non-cooperation and so on [1]. These key challenges are currently the focus of extensive research efforts in the P2P research community.

Search and resource location mechanisms are necessary for information reflection and collection, and are a fundamental and crucial building block of most P2P systems and determine, to a large degree, their efficiency and scalability. There are two types of P2P lookup services widely used for decentralized P2P systems: structured searching mechanism and unstructured searching mechanism. Those fundamental differences between the different types of P2P systems with regard to the implementation of the search mechanism indicate that not all P2P systems are compatible with AC.

Structured systems (e.g. Pastry [2]) are designed for applications running on well-organized networks, where availability and persistence can be guaranteed. These systems implement a distributed index or a distributed hash table (DHT) by using an efficient object-to-node mapping functionality with a hash function. The resulting cost of searching is logarithmic in the number of nodes in the system and therefore highly scalable. Since the location of objects is based on the exact knowledge of the corresponding name (or key), DHT-like algorithms lack flexibility and fault-tolerance. Even though more researches [3][4] have been made into developing more complex searches in systems, but these are also unsuitable for application in AC. They are unsuitable because users are widely distributed and come from non-cooperating organizations with highly dynamic joining and leaving.

Unstructured P2P networks, such as Gnutella for example, typically use blind search methods like flooding to distribute query messages throughout the network. This search method is very robust, flexible and easily supports partial-match and keyword queries. It's capability of adapting their behaviors dynamically can meet the changing specific needs of individual users in AC; and in addition will dramatically decrease the complexity and associated costs currently involved in the effective and reliable deployment of networks and communication services. However, the large volume of query traffic generated by the flooding of messages limits the scalability and efficiency of this approach. The most widely deployed unstructured P2P systems such as Kazaa (FastTrack Network) [5] and eDonkey (Overnet) [6] are hybrid systems which use a hierarchical approach to improve search efficiency. In hierarchical networks, nodes are partitioned into a set of logical groups, called clusters, by clustering algorithms. Each cluster can be treated as an abstract "super" node, and the resulting abstract nodes can be partitioned in the same way to form a higher-level cluster. However, the key problem is that most hierarchical structures use tree-based algorithms [7][8], layer 0 contains all peers, peers that become cluster heads at the $i^{th}$ level participate to the election of the $(i+1)^{th}$ level cluster heads [9]. Due to the single-parent nature, tree-based overlay is unbalanced and vulnerable to high "churn" (departure) rates of peers, since the departure of one peer affects all of its children.

In numerous P2P fashions, the individual nodes assign their own node identifiers. This makes establishing a level of trust extremely hard for initially unknown entities, since identities can easily be altered. Furthermore, it is possible for a malicious node to create multiple identities in order to gain control of a part of the system. This is referred to as a Sybil Attack [10]. The security of the solution mechanism [11] is based on the secure assignment of node identifiers via a trusted Certification Authority to stop potential Sybil Attacks. Such a trusted entity is typically not available in AC systems. On the contrary, in an open environment with mutually distrusting and potentially malicious participants, the lack of a trusted entity makes it extremely difficult to establish a level of trust mechanisms. And the growth of peer-to-peer services and the withering of centralized control make cooperative behavior essential to preventing free riding [12] and other selfish behaviors. A free rider is a node that has access to services and consumes resources without participating in any of them.

Incentive mechanisms penalizing free riders or rewarding users have been discussed [13][14][15], such as reputation mechanisms [16][17], monetary incentives [18][19] and game theory [20][21]. A big problem in proposing an incentive mechanism lies in the ability to track past behaviors in order to determine present service quality. Most of the existing work generally involves a trusted third party for records and query responses, which would suffer from the failure of a single point and in turn result in poor scalability. That is to say, guaranteeing the history of nodes' behaviors is an extremely challenging problem and an open research issue for AC. Thus, how to make the node ID unique and how to make the records of nodes' behavior consecutive in a highly dynamic and decentralized environment such as AC is a key question.

In the final analysis, to build an ideal P2P overlay for autonomic communication and guarantee the efficient incentive mechanism, there are three key problems to be solved: (1) how to maintain the overlay inexpensively in front of the dynamic topologies; (2) how information can be located and routed at a low cost; (3) how to make the overlay robust.

In this paper, we will present an efficient and low-cost P2P overlay called a Maya Pyramid Overlay (MPO) describing the above three key problems of P2P systems. The main idea of this overlay is to combined the merits of DHT P2P overlay (i.e. structured and efficient node structure and query mechanism) with that of an unstructured P2P overlay (i.e. scalability, adaptivity and robustness) to build a novel overlay, in which every node has an unique ID and a consecutive information exchange history even though the dynamic joining and leaving of nodes is without authentic global servers. In MPO, (1) update and backup only take place in two neighboring levels or layers. The maintenance cost is alleviated because the update and processing costs are lower. It differs from tree-based algorithms in which there are parents and grandpa nodes so that the update and backup information have to send to more than two levels. (2) Small world [22] characteristic which is usually utilized in structured DHT-like overlay is applied to build the unstructured overlay making the response mechanism in the overlay being typically the best way to answer the queries, and provides anonymity and censorship protection with efficiency. (3) A three information center (Tri-IC) mechanism is presented; each IC has a different duty and backup the other in order to lighten the loads of cluster heads in a traditional hierarchical overlay and thus preventing a single fault. The mechanism differs from most cluster head selection mechanisms only consider nodes' capability such as the end-to-end delay to other nodes [23] in one cluster or offered outbound bandwidth [24]. It also brings higher search efficiency and is more robust than the popular redundant mechanism.

## 2. The Organization of MPO

2.1 The formation of the overlay

In the overlay, nodes are divided into many clusters called autonomous system (AS). Inspired by the idea of the prime meridian or longitude 0 used to describe longitude and latitude, we propose using a normative node called the origin node (ON), in which ASs in the overlay calculate their approximate coordinates in the same virtual space . ON is not a permanent node since our overlay is a full self-organising architecture; it is mainly used to form the overlay through consideration of the physical situation. Moreover, ON disappears when the initial overlay becomes a stable overlay. When ASs can maintain the structured overlay by themselves, the overlay is called stable.

Our overlay consists of ASs, levels and layers. The maximum AS size is $d$+3. That is to say, the maximum peer number in one AS is $d$+3. The AS with the maximum peer number is called FULL AS (FAS). Before the overlay can become a stable overlay, nodes in one AS are sorted out according to their distances from the ON and the unique node ID is given. Nodes belonging to the same AS have an IP address sharing the longest common prefix. The ASs are positioned on a gradient which increases the farther away they are from the ON , and each AS has a unique AS ID. Many grouped ASs are constituent parts of a level, which has a threshold ($AN_i$), where $i$ is the number of the AS in the level, to the AS number. When one AS is a FAS, a new AS will be created as the next AS of the FAS in the same level. Let $L_{max}$ be the distance between the new node and the farthest node in an AS. If there are many candidate ASs, the new node will choose an AS whose $L_{max}$ is the minimum. Levels are sorted out according to their distance from the ON. A new level ($i+1$) will be created above the primary level $i$ to set new ASs while the number of ASs on the primary level reach the predefined threshold. Each level has a unique level ID. The planar space has many levels called layers, which are further increasing gradients guided by their distances from the ON. When the number of levels in a layer reaches the predefined threshold ($L_i$), where $i$ is the number of the layer, the structure grows to a three-dimensional space, i.e. a new $(i+1)^{th}$ layer will be created above the $i^{th}$ layer to set new levels. Thus new ASs will be initiated in a new layer above the primary layer. At the end of this process, an overlay-structure that matches the underlying network can be constructed based on the clusters of nodes that are close together and in proximity to each other.

Fig. 1 is our logical overlay, where $H$ is the number of levels and the number of layers. Every layer includes many levels and ASs communicating to each other by cluster headers called information centers (ICs) appearing in all ASs, levels and layers. The predefined bound $d$=2 in Fig.1. The red, black and blue dots denote $IC_{level}$, $IC_{layer}$ and $IC_{local}$ respectively, which will be introduced in section 2.3. We use the tri-IC mechanism not a solo cluster head with redundancy so that the overlay is called Maya pyramid overlay (MPO).

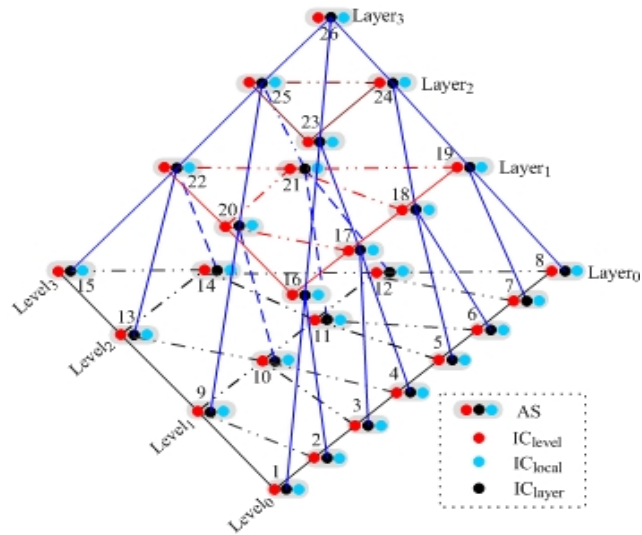

Fig.1 The perspective of the MPO structure (H=4, d=2)

The edge between neighbor ASs (e.g. the edge between AS 9 and AS 20; or the edge between AS 9 and AS 2) is one of the *long-range contacts*, and the edge between any two nodes in the same AS is one of the *local contacts*. L*ong-range contacts* and *local contacts* are presented in [22] to build small world networks, while our *contacts* are not randomly chosen. When the overlay-structure is a stable overlay, the ON ceases to exist. Let $D_{max}$ be the average distance between the new node and the ICs in an AS. If there are many candidate ASs, the new node will choose an AS whose $D_{max}$ is the minimum. ICs in our work cooperate with each other to provide this geographical partitioning feature. Therefore the problem of fixed landmarks can be avoided since the unavailability of ICs will affect only that local region.

2.2 Source ranking

Generally P2P systems consist of a large number of nodes. Nodes can enter and depart any time. Moreover, node behavior cannot be trusted. There is always the possibility of free riders in the system who selfishly download the content without treating the others fairly. We have focused on the overall system performance and not on individual gains. Consequently the contributing nodes suffer because of these free riders. This also has significant impact on the upload utilization and fairness ensured by the preferential and strata based scheme.

In our overlay-structure, we present a mechanism called Source Ranking where a peers' group is formed by the online queuing function *SR*. When the information exchange begins between peer $P_i$ and $P_j$, it is assumed that query is sent from $P_i$ to $P_j$. Then $E(P_i, P_j)$ denotes the $P_i$'s evaluation to $P_j$ about the completion of the query.

$$E(P_i, P_j) = \sum_{\forall q \text{ answered by } P_j} \text{Qsim}(q_j, q)^\alpha * N(P_i, P_j) / T(P_i, P_j) \qquad (1)$$

Parameter $\alpha$ improves the power of similarity of queries. Since $\alpha$ is bigger, queries that are more satisfied are given a higher evaluation. That is, the more similar queries $P_j$ completes, the larger evaluation $P_i$ gives. In order to find the most likely peers to answer a given query we need a function Qsim: $Q^2 \rightarrow [0, 1]$ (where Q is the query space), to compute the similarity between different queries. $N(P_i, P_j)$ indicates the communication times between peer $P_i$ and $P_j$. $T(P_i, P_j)$ denotes the time of information exchange between peer $P_i$ and $P_j$. When the similarity between the different queries is the same, we can assume that the peer has more communication times in unit information exchange rather than in time joining the AS in order to decrease the percentage of the free riders.

The cosine similarity (formula 2) metric between 2 vectors ($\vec{q}$ and $\vec{q}_i$) has been used extensively in information retrieval, and we use this distance function in our setting. Let $L$ be the set of all requirements for resources or services that have appeared in queries. We define an $|L|$-dimensional space where each query is a vector. For example, if the set *L* is the requirements {A, B, C, D} and we have a query A, B, then the vector that corresponds to this query is (1, 1, 0, 0). Similarly, the vector that corresponds to query B, C is (0, 1, 1, 0). In the cosine similarity model, the similarity *sim* of the two queries is simply the cosine of the angle between the two vectors.

$$sim(q, q_i) = \cos(q, q_i) = \frac{\sum (\vec{q} * \vec{q}_i)}{\sqrt{\sum (\vec{q})^2} * \sqrt{\sum (\vec{q}_i)^2}} \qquad (2)$$

Thus, the source ranking of $P_j$ is:

$$SR(P_j) = \sum_{\forall i \neq j \in \{\text{the same AS}\}} (E(P_i, P_j) * D(P_i, P_j)) \Big/ \sum_{\forall i \neq j \in \{\text{the same AS}\}} D(P_i, P_j) \qquad (3)$$

Where $D(P_i, P_j)$ is the inverse proportion function of the distance between peer $P_i$ and $P_j$. The closer peers lie in geography, the bigger the function $D$ is. It improves the power of peers adjacent geographically in order to avoid influence of network congestion.

2.3 Tri-IC mechanism

Each node has the unique 128 bit node identifier in our system. The node ID is used to indicate a node's position in the overlay, which ranges from 0 to $2^{128}$ -1. The node ID is defined by the layer number, the level number, the AS number and the node number immediately after a node joins the system. The letters *i, j, k* and *t* respectively denote the layer number, the level number, the AS number and the node number, which are shown as follows:

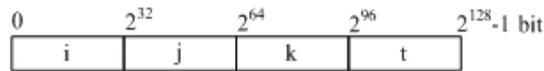

Fig.2  The structure of node ID

$N_{k,t}^{i,j}$ denotes the $t^{th}$ NN $t \in \{1 ... d\}$ of the $k^{th}$ AS at the $j^{th}$ level in the $i^{th}$ layer. It is obvious that the node ID is global unique.

There are three ICs called $IC_{local}$, $IC_{level}$ and $IC_{layer}$ in each AS, and the three nodes are connected to each other. The other nodes in the AS are called normal nodes (NN) and their applications and resources are abstract as the information. NNs don't have to keep information tables of each other but only information about the IP and ID of the $IC_{local}$, $IC_{level}$ and $IC_{layer}$. In our paper we will define the maximum number of NNs as $d$. The tri-IC are the nodes which have the top three biggest values of *SR* in ASs. The ICs have different functions but contain the same backup. $NN_i^j$ denotes the $j^{th}$ NN and $IC_\sigma^i$ denotes the $IC_\sigma$ ($\sigma \in \{local, level, layer\}$) in AS $i$ ($i \in [1,26]$) in Fig. 1.

**$IC_{local}$**: $IC_{local}$ is connected by and receives information such as ID, IP, applications, resources, activity histories (i.e. the values of *SRs*) and status from most $d$ local NNs. For example, $NN^i$ directly connect to and store their information in $IC_{local}$ $IC_{local}^i$. $IC_{local}$ also keeps the backup of information and sends its own information (i.e. information of NNs) to local $IC_{level}$ and local $IC_{layer}$.

**$IC_{level}$**: One $IC_{level}$ in the $i^{th}$ level communicates with at most $d$ the $(i-1)^{th}$ $IC_{level}$s. Each $IC_{level}$ is connected with only one $IC_{level}$ in the upper neighbor level. A local $IC_{level}$ submits the application and resource information of its AS (i.e. information from local NNs aggregated in local $IC_{local}$ are backed up to $IC_{level}$) to the corresponding $IC_{level}$ in the upper neighbor level. E.g. $IC_{level}^9$ and $IC_{level}^{10}$ are two $IC_{level}$s connected to the upper neighbor level $IC_{level}$ $IC_{level}^{13}$. $IC_{level}^{13}$ backs up information in $IC_{local}^{13}$ and $IC_{layer}^{13}$, and sends information about local NNs (e.g. $NN_{13}^j$) to its upper neighbor level $IC_{level}^{15}$. To recover from all ICs' crash, $IC_{level}$ also stores upper neighbor level NNs' IP addresses.

**$IC_{layer}$**: One $IC_{layer}$ in the $i^{th}$ layer communicates with at most $d$ $IC_{layer}$s in the $(i-1)^{th}$ layer while the AS which lies in the highest level has 1 extra subordinate from the highest level AS in the lower neighbor layer. Each $IC_{layer}$ is connected with only one

IC$_{layer}$ in the upper neighbor layer. Because of characteristics of our overlay, IC$_{layer}$ in the upper layer only backs up the ID and IP information and activity histories of the nodes of its lower neighbor layer ASs. For example, $IC_{layer}^{20}$ stores the information such as ID, IP and values of *SRs* receives from $IC_{layer}^{9}$, including those information of $IC_{local}^{9}$, $IC_{level}^{9}$, and $NN_{9}^{j}$ etc. Since the distance across layers is longer than the distance between levels in the same layer, the possibility of a fault in transferring large data increases. However, the amount of ID, IP and activity history information is small and it is suitable to be transferred across layers in order to improve the whole protocol's robustness. The IC$_{layer}$ also keeps the backup of information and sends its own information (i.e. ID, IP and value of *SR* about lower neighbor layer nodes) to local IC$_{level}$ and local IC$_{local}$. To recover from a tri-IC's crash, the IC$_{layer}$ also stores its upper neighbor layer NNs' IP addresses.

Our information exchanges are triggered by a trigger mechanism called burst trigger which is triggered by nodes' status change. In our overlay, each node including ICs and NNs has a three status function: free, busy and normal. The information exchange between two nodes is driven by the node whose status is changed into the free one from one of the other two kinds of statuses, but it is not updated periodically. Judging from the condition described above, the load in IC$_{level}$ is the heaviest. So we choose the node with the highest value of *SR* in an AS as the IC$_{level}$ in this AS. Then the nodes having the ordinal highest (the 2$^{nd}$ highest and the 3$^{rd}$ highest) values of *SR* are named as IC$_{local}$ and IC$_{layer}$ respectively. The other nodes in this AS are NNs which don't have to keep information tables of each other but only information about the IP and ID of the IC$_{local}$, IC$_{level}$ and IC$_{layer}$.

The most difference between MPO and traditional hierarchical P2P overlay [9][30] is that each $IC_{level}$ does not back up the information of the next level – its lower neighbor level $IC_{level}$. It is similar to each $IC_{layer}$. For instance, $IC_{level}^{15}$ does not receive the information from $IC_{level}^{9}$ or $IC_{level}^{10}$, but only the information from $IC_{level}^{13}$ and $IC_{level}^{14}$. That is to say, the backup information is only kept in two consecutive levels, not recursively stored in the next level of the upper neighbor level as in most hierarchical structures. The latter condition may result in the upper level being more important and storing more information. Therefore, the upper level ICs may be more vulnerable. It is the same with layers.

## 3. The maintenance of MPO

In this section, we propose the maintenance of the overlay, including joining, leaving and the mechanism for preventing free riders. Peers in level *j<H-1* are partitioned into ASs of size in [$\delta$, *m*] (The analysis assumed that each AS are connected by at least *k* lower neighbor ASs. $m > \delta \geq 1$ are two constant).

**Lemma 1:** it is assumed that *k* is a constant ($k \geq 2$), to $\forall x > 0$, $k^x \geq kx$.

Proof: Here we use epagoge.

(ⅰ) To *x*=1, k=*k* is true.

(ⅱ) To *x*=2, since $k \geq 2$, $k^2 - 2k = (k-1)^2 - 1 \geq 0$, thus $k^2 \geq 2k$,

(iii) to *x>2*, it is assumed that $k^{x-1} \geq k(x-1)$, then,

$$k^x = k \cdot k^{x-1} \geq k \cdot k(x-1) \quad (4)$$

Since $k \geq 2$, $1 + \frac{1}{k-1} \leq 2 < x$, that is,

$$\frac{k}{x} < k - 1 \quad (5)$$

From inequation (5) we conclude

$$k^2(x-1) > kx \quad (6)$$

Combined inequation (4) and (6), $k^x \geq kx$, by the epagoge, lemma 1 is proven.

Lemma 2: it is assumed that k is a constant and $k \geq 2$, then $k^x > (k-1)x - k$ (x>0).

Proof: since $k \geq 2$ and x >0, then $(k-1)x - k < kx - k < kx$, from lemma 1, lemma 2 is proven.

Theorem 1: If one $IC_{level}$ in the ith level communicates with at most d the (i-1)$^{th}$ $IC_{level}$s, and one $IC_{layer}$ in the i$^{th}$ layer communicates with at most d the (i-1)$^{th}$ $IC_{layer}$s. The height $H$ of levels or the layers in the overlay is less than $\log_d M + 1$, where $M$ is the number of ASs.

Proof: To level H-2, the group number is at most $d$, Thus, to layer 0 (the number of levels is H-1), the number of all ASs is

$$\sum_{n=1}^{H} d^{n-1} = \frac{1 - d^H}{1 - d}$$

Recursively, to layer 1, the number of levels is H-2, thus the number of all ASs is

$$\sum_{n=1}^{H-1} d^{n-1} = \frac{1 - d^{H-1}}{1 - d}$$

Then, to the overlay, the number of ASs is

$$\frac{(1-d)+(1-d^2)+(1-d^3)+\ldots+(1-d^H)}{1-d} = \frac{H - d \cdot (1+d+\ldots+d^{H-1})}{1-d} = \frac{H - d \cdot \frac{1-d^H}{1-d}}{1-d} = \frac{H - d \cdot H - d + d^{H+1}}{(1-d)^2}$$

Then, $\frac{H - d \cdot H - d + d^{H+1}}{(1-d)^2} = M$

Combined lemma 2, we have

$$M = \frac{H - d \cdot H - d + d^{H+1}}{(1-d)^2} > \frac{d^{H+1} - d^H}{(1-d)^2} = \frac{d^H}{d-1} > d^{H-1} \quad (7)$$

From inequation (7), we have $H < \log_d M + 1$. Theorem 1 is proven.

Theorem 2: The worst-cast node degree of the multicast overlay is at most *d+4*.

Proof: Consider a node *X* in one AS. There are four possibilities:

(1) X is the NN: X only links to $IC_{local}$ in the AS. Therefore, the degree of X is 1.

(2) X is the $IC_{level}$: An AS has at most d lower neighbor level ASs and only at most 1 upper neighbor level ASs, thus X has at most d+1 neighbor $IC_{level}$s (the degree is at most d+3). An exception holds for the highest level where the $IC_{level}$ has no upper neighbor level $IC_{level}$ (the degree is at most d+2)

(3) X is the $IC_{local}$: Each $IC_{local}$ is linked by at most d NNs, thus the degree of X is at most d+2.

(4) X is the $IC_{layer}$: An AS has at most d lower neighbor layer ASs and only at most 1 upper neighbor layer ASs. Thus, X's degree is d+3. The AS which lies in the highest level has 1 extra subordinate from the highest level AS in the lower neighbor layer (the degree is at most d+4). An exception holds for the highest layer where the $IC_{layer}$ has no upper neighbor layer $IC_{layer}$ (the degree is at most d+3).

In any case, the degree of a node can not exceed d+4, thus proving the theorem true.

Theorems 1 and theorem 2 summarize two properties any P2P structure, such as a tree, should desire. As clients join and leave, we must be able to adjust the 3D overlay without violating the adaptive rules. Overheads incurred by this adjustment should be kept small to keep the system scalable.

3.1 Node joining

There are many researches on bootstrapping. In this paper, we do not concentrate on how to bootstrap in our protocol; we just assume that every new node can always satisfy the following condition: For each new joiner, at least one close node existing in the overlay-structure can be found. This assumption allows the new node to join the AS close to itself. It is rational. For example, if the peers in City *A* are collected in one AS, there is a greater possibility for a new joiner in City *A* to meet one of the peers in the same city and to become a member of that AS.

- Suppose the node is a new node which has never joined the system before. If the bootstrapping node X is an $IC_{local}$, the new node submits its information to $IC_{local}$, and obtains a unique ID from the AS that is also a unique global ID. If the bootstrapping node X is not an $IC_{local}$, through the IP stored in X, the new node invites X's local $IC_{local}$ to join X's AS and submit all its information to local $IC_{local}$. Then the new node obtains a unique ID from the AS that must be a unique global ID.

- Suppose that the node had joined the system before. It rejoins the system through its record of the AS. If, upon accepting the new connection, the total number of nodes still is within the preconfigured bound $d+3$, the connection is automatically accepted. Otherwise, the AS must check if it can find an appropriate existing node to drop and replace the new connection.

In our system, ID is very important because it describes the identity and the situation of each node in the protocol. The AS has two characteristics as follows: (1) Each AS is globally unique. Nodes can join and leave system randomly, and they can also change the AS they belong to. However, each AS belongs to a fixed level of a fixed layer. (2) The scale of each AS is fixed. We predefine that the maximum number of one AS is

*d+3*. The node ID is globally unique but not lifelong. To make the structure adaptive, as we have mentioned, the scale of each AS is fixed.

Let $D_{avg}(x,r)$ be the average distance between the new node *x* and the ICs in an AS *r*. $D_i$ denotes the distance between node *x* and the node *i* in the overlay. $R_x$ is the maximum delay that this node *x* can be tolerable or acceptable. The joining algorithm is shown in Fig. 3.

```
procedure Joining
  S ← {r|CP_r = C_x ∧ (∃i ∈ r : D_i < R_x)}         if  min_r+1 ≤ max_r    /* have room*/
  q ← null                                              q ← r
  S' ← S                                            else
  While ( S' ≠ ∅ ){                                   if ∃i ∈ r : SR_i < min_EF
    r_j ← ⟨pick a random AS from S'⟩                     drop node i
    Calculate  D_avg(x, r_j)                             q ← r
    S' = S' − {r_j}                                   else
  }                                                   if ∃i ∈ r : max(D_avg(i,r)) > D_avg(x,r)
  While ( S ≠ ∅ ∧ q = null ){                           drop node i
    r ← {select (min(D_avg(x, r_j))∀r_j ∈ S)}           q ← r
    S = S − {r}                                     }
```

Fig. 3 The pseudocode of joining algorithm

If the nearest AS to the new joiners is a FAS how can we keep the initial structure? Our method firstly drops the free riders from the system, and then substitutes the nodes with longer distances. Each AS has a minimum value $Min(Ef)_k^{i,j}$ of peers' *SR*, where *i*, *j*, *k* denotes the AS's layer number, level number and its location in the level respectively. $Min(Ef)_k^{i,j}$ is dynamically decided by a number of nodes and the loads in an AS. If one node's value of *SR* is less than $Min(Ef)_k^{i,j}$, we consider it as a free rider. The new node *x* chooses a group in which all of the nodes are within the radius $R_x$ centered at itself, and where the nodes share the longest common prefix (CP) of IP addresses with the new node *x* are aggregated. Therefore, if a new node intends to join a FAS, this FAS will firstly drop the free riders. If the AS has not a free rider, node *x* has to compare its distance with the node with the longest distance in the AS. (1) If the joiner's distance is longer, it will join a neighbor group which is the second nearest to it; (2) if the joiner has a shorter distance, the old node is substituted by the new joiner and the old one joins a neighbor group which is the second nearest to it.

Many researchers have presented the node ID [25][26]. However, they do not deal with the mechanisms of how to guarantee the uniqueness of IDs in a system, and the stability of the system. In this paper, we propose an entirely automatic scheme: each AS only knows the global unique IDs of several neighboring ASs.

In ASs communicating with each other by Peer-to-Peer without a global IC, it is important to prevent nodes from registering in several ASs so that one node has several IDs. Our solution to this is to define a threshold distance ($T\_dist$) between neighboring ASs in the same level as follows:

$$T\_dist = \frac{\sum_{\sigma \in \{IC\}} (p_\sigma^{cri} * Dist_\sigma)}{\sum_{\sigma \in \{IC\}} p_\sigma^{cri}} \qquad (8)$$

Where $\sigma$ is one type of the three ICs, $Dist_\sigma$ denotes the distance between two neighboring $IC_\sigma$ at the same level. E.g., $Dist_{local}$ denotes the distance between $IC_{local}$s in two neighbor AS at the same level. $p_\sigma^{cri}$ denotes a parameter defined by the position of $IC_\sigma$ in its AS. It should be noted that ICs are changed by the dynamic joining and leaving, Thus, $p_\sigma^{cri}$ is necessary because the position of each IC is dynamic. The discuss of $p_\sigma^{cri}$ could be found in our other paper [27]. In our overlay, the unique ID not only identifies the peer, but is also used to determine the peer's location in the overlay so that every peer is capable of supervising each other via the ID to prevent cheating. Both the 3D overlay and the unique ID mechanism provide secure cooperation among peers. Whenever and whatever peer joins or leaves the P2P system, they are always restricted in a limited location in the overlay. Thus, the activity history of any peer can always be found in remained peers in the same AS.

From the above theoretical analysis, we conclude that in our overlay a node can not register in more than one AS. The nodes must join the nearest AS. Therefore, if a new node wants to join an AS farther than *T_dist*, the AS will regard it as whitewasher so that the join request must be refused, and the free riders are always penalized in functions of the P2P system. The detailed game theory about the behavior of nodes will be introduced in another paper.

3.2 Node leaving

Normal leaving

There are two ways that the nodes can normally leave the system: ICs (i.e. $IC_{local}$, $IC_{level}$ and $IC_{layer}$) leave; and NNs leave.

1. Normal leaving of ICs. In our system, only one IC can normally leave at one time. If all the three ICs want to leave the system, they have to leave one by one following a principle called the first application first leaving (FAFL). While an IC is preparing to leave the system, it uses the backup information and chooses the NN that has the highest value of *SR* among NNs as its substitute.

(1) If $IC_{level}$ at the $j^{th}$ level is the leaving node, it should broadcast the information of its substitute to (a) local $IC_{local}$ and local $IC_{layer}$; (b) $d$ the $(j-1)^{th}$ level neighbor $IC_{level}$s; (c) and to the $(j+1)^{th}$ level neighbor $IC_{level}$.

(2) If $IC_{local}$ is the leaving node, it should broadcast information of its substitute (a) to the local $IC_{level}$ and $IC_{layer}$; (b) and to $d$ local NNs.

(3) If IC$_{layer}$ in the $i^{th}$ layer is the leaving node, it should (a) broadcast its information to local IC$_{local}$ and IC$_{level}$; (b) and it should broadcast information to the $(i+1)^{th}$ layer IC$_{layer}$ and to $d$ the $(i-1)^{th}$ layer neighbor IC$_{layer}$s.

2. Normal leaving of NNs. It is simple for NNs to leave because the only thing that the node has to do is to submit its leaving request to a local IC$_{local}$ and change its status in the local IC$_{local}$ to be 'offline'.

Abnormal leaving

There are also two ways that the nodes abnormally leave the system: ICs leave and NNs leave. In our system, one IC sends life signals to the other corresponding ICs periodically. If, for a long time an IC does not receive life signals from another IC, this means that it has abnormally left the system.

1. ICs abnormal leaving. There are two situations as follows:

(1) The tri-IC does not leave the system simultaneously. The remaining IC will choose the node which has the highest value of *SR* as the new IC, and its backup information will be copied to the new IC. Then the new IC broadcasts its information to the corresponding ICs in the system.

(2) All tri-ICs leave the system abnormally. The upper level neighbor IC$_{level}$ chooses the node *x* in its lower neighbor level AS with the highest value of *SR* as the new IC$_{level}$, and copies the backup information to the new IC$_{level}$. Then the new IC$_{level}$ chooses the new IC$_{local}$ and IC$_{layer}$ following the method described above. The tri-IC broadcast their information to the corresponding ICs in the system as described in section 2.3. For example, as shown in Fig. 1, if $IC_\sigma^i$ denotes one of ICs in AS *i* where $\sigma \in \{local, level, layer\}$, and $NN^i$ denotes one of NNs in AS *i*. When all ICs in AS 20 crashed, (a) $IC_{level}^{22}$ chooses one $NN^{20}$ whose *SR* value is the highest to be the new $IC_{level}^{20}$; (b) $IC_{level}^{20}$ receives information about AS 20 backed up in $IC_{level}^{22}$ from the latter; (c) $IC_{level}^{20}$ chooses two $NN^{20}s$ whose *SR* value is the ordinal highest to be the new $IC_{local}^{20}$ and $IC_{layer}^{20}$ respectively; (d) $IC_{local}^{20}$ receives information about $NN^{20}s$ from $IC_{level}^{20}$ and connect with $NN^{20}s$; (e) Simultaneously, $IC_{level}^{16}$ and $IC_{level}^{17}$ reconnected to $IC_{level}^{20}$ through inviting any $NN^{20}$ to get the address of the new $IC_{level}^{20}$; $IC_{layer}^{9}$, $IC_{layer}^{10}$ and $IC_{layer}^{25}$ reconnected to $IC_{layer}^{20}$ through inviting any $NN^{20}$.

2. NNs abnormal leaving. The necessary information such as ID, IP and activity history are still kept in ICs to distinguish them from the nodes which have never joined the system. In the worst situation, several ASs will break down. Then the new nodes will have to reuse the formation process to recreate ASs based on the existing AS.

**4. Evaluation**

4.1 Modeling and methodology

It is impossible to model all the dynamics of an Internet-based P2P system. In this paper, we are not trying to resolve small quantitative disparities between different algorithms, but instead are trying to reveal fundamental qualitative differences. While our simple models do not capture all aspects of reality, we hope they capture the essential features needed to understand the qualitative differences. To evaluate the performances of unstructured P2P overlay, we look at three aspects of a P2P system: P2P network topology, query distribution and replication. By network topology, we mean the graph formed by the P2P overlay network; each P2P member has a certain number of "neighbors" and the set of neighbor connections forms the P2P overlay network. By query distribution, we mean the distribution of frequency of queries for individual files. By replication, we mean the number of nodes that have a particular file. During our study of search algorithms we assume static replication distributions.

- Random Topology with Power-Law characteristic (RTPL): This is a random graph with different scales. The node degrees follow a power-law distribution: if one ranks all nodes from the most connected to the least connected, then the $i^{th}$ most connected node has $\omega/i^{\alpha}$ neighbors, where ω is a constant. Many real-life P2P networks including real Internet network have topologies that are power-law random graphs [29].

- Super-node topology: We ran simulations using a standard super-node topology [31]. It is a two-level hierarchy, consisting of a first level of interconnected peers called super-peers and a second level of so-called leaf nodes or normal peers, which are only connected to a single super-peer. In super-node topology, searches are flooded among super-peers. In the paper, the term "node" is used interchangeably with "peer". Each super-node has two backup nodes. The peers including super-node, its backups and normal peers form a cluster. From well-known observations on Gnutella, it is observed that even powerful nodes maintain only about 10 neighbors [14]. We set the total number ($c_{size}$) in any cluster is $c_{size} \in [5, 15]$.

- Square-root topology [31]: Consider a peer-to-peer network with $N$ peers. Each peer $k$ in the network has degree $d_k$ (that is, $d_k$ is the number of neighbors that $k$ has). The total degree in the network is $D$, where $D = \sum_{k=1}^{N} d_k$. Each peer $k$ maintains two counters: $Q_{total}^k$, the total number of queries seen by $k$, and $Q_{match}^k$, the number of queries that match $k$'s content. $g_k$ denotes the proportion of searches submitted to the system that are satisfied by content at peer $k$, that is, $g_k = Q_{match}^k / Q_{total}^k$, then a square-root topology has $d_k \propto \sqrt{g_k}$ for all *k*. To construct a square-root topology, when peers join the network, they make random connections to some number of other peers. The number of initial connections that peer *k* makes is denoted $d_k^0$. Then, as peer *k* is processing queries, it gathers information about the popularity of its content. From this information, peer *k* calculates its first estimate of its ideal degree, $d_k^1$. If the ideal degree $d_k^1$ is more than $d_k^0$, peer *k* adds $d_k^1 - d_k^0$ connections, and if the ideal degree is less than $d_k^0$, peer *k* drops $d_k^0 - d_k^1$ connections. Over time, peer *k* continues to track the popularity of its content, and re-computes its ideal degree ($d_k^2, d_k^3 ...$). Whenever its ideal degree estimate is different from its actual degree, peer *k* adds or drops connections.

- MPO

Though there are many other unstructured searching algorithms, such as Iterative deepening [32], Biased high degree [33], Most results and Fewest result hops [34] etc., the flooding search method is very robust, flexible and easily supports partial-match and keyword queries. [35] analytically demonstrates that random walks are useful to locate popular content when the topology forms a super-peer network, and have better performance in topologies with power-law characteristic. Moreover, it is proofed in [31] that a square-root topology is optimal for random walk searches. Considering all four topologies being compared in our paper, it is rational that we just use this two search algorithms to evaluate the performance of topologies.

- Flooding: When a peer receives a search message, it both processes the message and forwards it to all of its neighbors in the overlay network. Each message is given a time-to-live value *ttl*, and search messages get flooded to every node within *ttl* hops of the source. There are two flooding algorithms: 1) unrepeated flooding, in each query, the query message will not send to a node which has received that message; 2) repeated flooding, message can be sent to the same node repeatedly.

- Random walk: When a peer receives a search message, it processes the message and then forwards it to one or several randomly chosen neighbors (called walks). Messages continue random walking until either a predefined number of results are found (again, predefined by the user), or a *ttl* is reached. Random walk *ttl* values are high and exist mainly to prevent searches from walking forever. In the paper, walks=4.

Studies have shown that Gnutella, Media and Web queries tend to follow Zipf-like distributions [36]. Thus, in our simulations, the number of each file follows a Zipf-like distribution according to its popular degree. It is assumed that there are *m* original files. And $q_i$ represents the relative popularity, in terms of the number of queries issued for it, of the $i^{th}$ object. Then, we can get $q_i \propto 1/i^\alpha$, where each file $f_i$ is replicated on $r_i$ nodes, and the total number of files stored in the network is $R$ and $\sum_{i=1}^{m} r_i = R$. In our simulations, the replication of a file $f_i$ is proportional to the query probability of the file. If one assumes that only nodes requesting a file store the file, then the replication distribution is usually proportional to query distribution (i.e. $r_i \propto q_i$). In fact, the trends of performance results among topologies are the same even with different parameters in Zipf-like distributions, we just show the results when *α*=0.726 and *m*=300.

4.2 Simulations

We conducted our experiments on three types (i.e. RTPL, super-node topology and MPO) of physical network topologies with different node-number (i.e. $N$=500, 1000, 1500 and 2000) generated by using the GT-TTM library [28]. For constructing square-root topology, we choose a maximum degree $d_{max}$, representing the degree we want for a peer whose popularity $g_k = 1$. Of course, it is unlikely that any peer will have content matching all queries, so the actual largest degree will almost certainly be less than $d_{max}$. Then, we can define $D$ as $D = d_{max} \cdot \sum_{i=1}^{N} \sqrt{g_i}$. If the popularity of a peer's content is very low, then $d_k$ will be very small. If peer degrees are too small, the network can become partitioned, which will prevent content at some peers from being found at all. In the worst case, because $d_k$ must be an integer, so the ideal degree might be zero. Therefore, we define a value $d_{min}$, which is the minimum degree a peer will have. The degree a peer will aim for is:

$$d_k = \begin{cases} round(d_{max} \cdot \sqrt{Q_{match}^k / Q_{total}^k}) & \text{if greater than } d_{min} \\ d_{min} & \text{otherwise} \end{cases} \quad (9)$$

The square root constructing progress can be summarized as follows [31]: 1) We choose a maximum degree $d_{max}$ and minimum degree $d_{min}$, and fix them as part of the peer-to-peer protocol. 2) Peer $k$ joins, and makes some number $d_k^0$ of initial connections ($d_{min} \leq d_k^0 \leq d_{max}$). 3) Peer $k$ tracks $Q_{match}^k$ and $Q_{total}^k$, and continually computes $d_k$ according to equation (9). 4) When the computed $d_k$ differs from peer $k$'s actual degree, $k$ adds or drops connections. We ran 10000-time simulations to measure the performance of searches over time as the topology adapted under the square root constructing progress, the parameters for the square-root-construct algorithm are shown in Table 1.

Table 1 Parameters for constructing square-root topology

| Parameter | Value | | | |
|---|---|---|---|---|
| | $N$=500 | $N$=1000 | $N$=1500 | $N$=2000 |
| $d_{max}$ | 40 | 80 | 100 | 160 |
| $d_{min}$ | 3 | 3 | 3 | 3 |
| $d_k^0$ | 4 | 4 | 4 | 4 |

We experimented with several parameter settings, and found that these settings worked well in practice.

### 4.2.1 Scalability

For simulations in this section, we generate every type of topologies 100 times respectively under various system scales. Node degree information of the four graphs under various node numbers are shown in Fig.4. The number of layers and levels (which are defined in [37]) in MPO are both $H$=3.

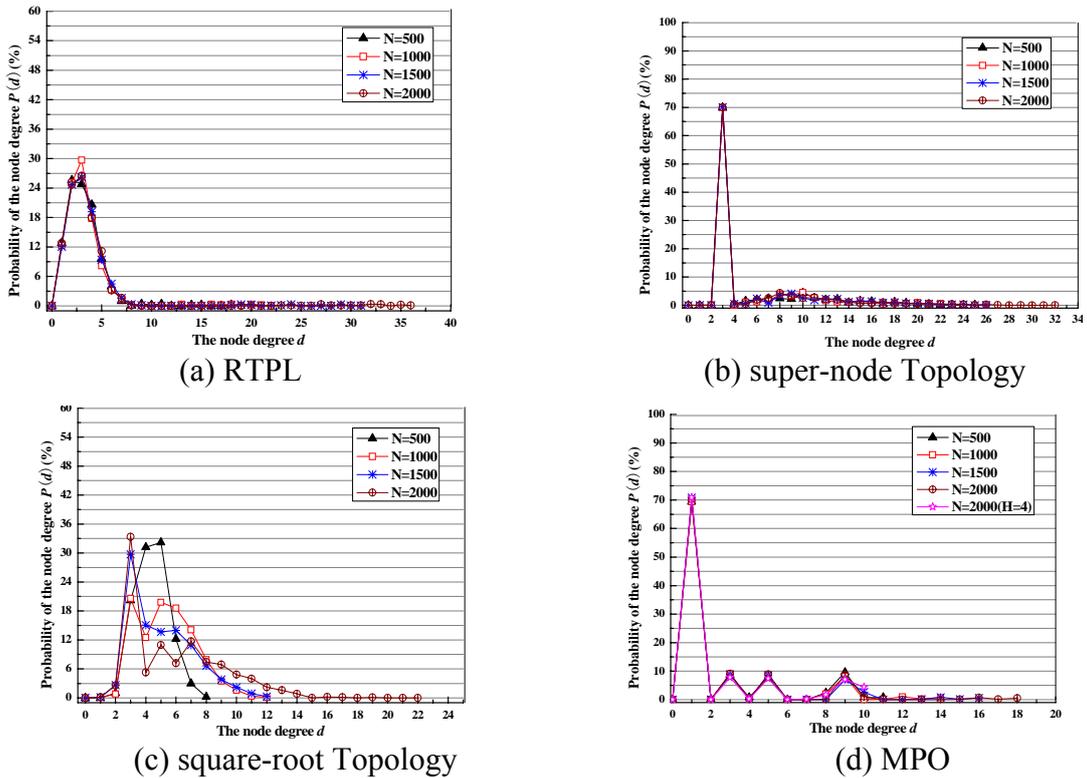

(a) RTPL  (b) super-node Topology

(c) square-root Topology  (d) MPO

Fig. 4. Distribution of node degrees in the four network topology graph

From Fig.4, we can see that the value of the most degree node's degree respectively in RTPL, super-node topology and square-root topology greatly increases as the system's scale increases. For example, in RTPL, the value of the node with the most degree is 18 when $N$=500, while that is 36 when $N$=2000. In super-node topology, the value of the node with the most degree is 25 when $N$=500, while that is 32 when $N$=2000. In square-root topology, the former is 8 and the latter is 22. However, in MPO, the maximum degrees change from 11 to 18 when the N changes from 500 to 2000, that is, the value of the node with the most degree does not change quickly (i.e. the change is limited in unit position). Moreover from Fig.4, it is clear that the node's degrees in MPO almost do not change as the system scale increases, or the number of layers increases (e.g. $H$=4). Fig.4 shows that we are able to adjust such a 3D structure as MPO without violating the adaptive rules as clients join and leave. Overheads incurred by this adjustment could be kept small to keep the system scalable.

4.2.2 Query success rate

We evaluate the query success rate of four topologies with different system scales by the popular searching algorithms introduced in section II. In the simulation, there are total 4162 files including both original files and replications. For each experiment, the source is chosen randomly while the requested file is chosen according to Zipf distribution. We report the mean values of results obtained through 120000 runs.

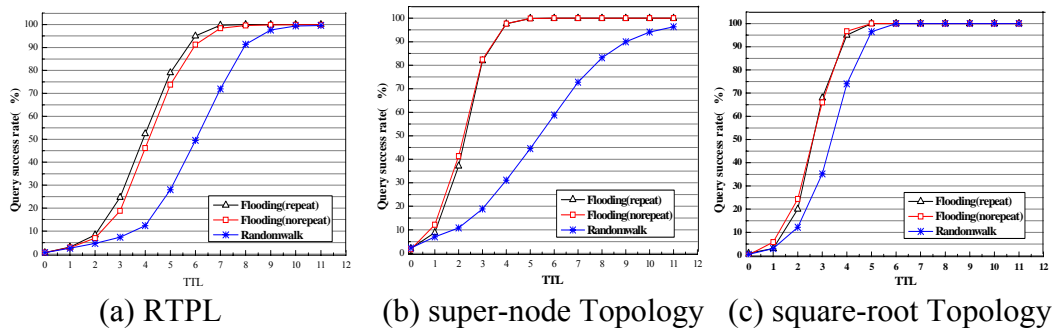

(a) RTPL      (b) super-node Topology   (c) square-root Topology

Fig. 5. Query success rate for different searching algorithms in different topology

From Fig. 5, we find that whatever the system scale and topologies are, the query success rate when using repeated flooding algorithm is the highest, though it is very near to that when using unrepeated flooding algorithm, especially in super-node

MPO: An Efficient and Low-cost Peer-to-Peer Overlay for Autonomic Communications

topology and square-root topology. On the contrary, the rate using random walk algorithm is the lowest. So we just use one middle algorithm (i.e. unrepeated flooding) to evaluate and compare the query success rate in all four topologies when $N$=2000, as shown in Fig. 6.

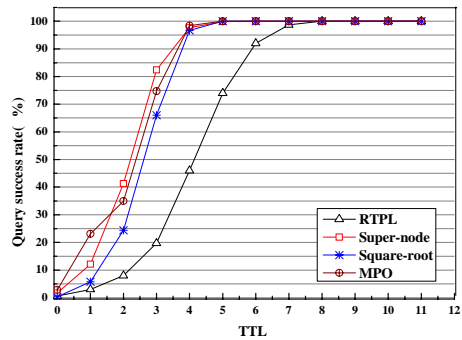

Fig. 6. Query success rate for various topologies

From Fig.6, the query success rate in MPO is the highest when TTL is no more than 1, that is, the satisfied query in one hop in MPO is more than others. We think that it because of the 3D structure so that one AS has more neighbor-ASs.

4.2.3 Cost and load balancing

In such high dynamic environments as P2P systems, the requirement for system stability is more important than for query success rate. By simulating, we find that there is the same regularity for the cost line distribution of each searching algorithm whichever the topology is. Therefore, we compare the four topology cost using unrepeated flooding algorithm when $N$=2000; the results are shown in Fig.7. The $x$-axis shows the upper bound of hops permitted by every searching, the $y$-axis shows the average system cost.

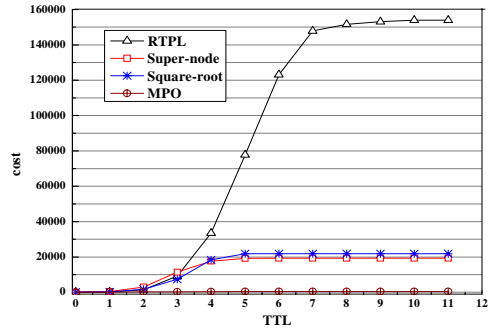

Fig. 7. The searching cost in various topologies

In Fig.6, it is clear that the query success rates of super-node topology, square-root topology and MPO are all near or equal 100% when *TTL*=4. However, we can find in Fig.7 that the searching cost in MPO is much less than that in super-node and square-root topology. And the cost in MPO does not remarkably increase as the searching hops increasing, on the contrary, the costs in other 3 topologies rapidly increase after the first hop. Using different searching algorithms in various topologies, the situation of the disturbing to each node is different. The average disturbing times are more, the load of the system is heavier. Thus, the possibility of system crash is higher. Therefore, a good topology should have small disturbing rate in each search for guaranteeing the stability of the system. Just evaluating the total system cost maybe cover up the load unbalancing that some peers are disturbed too much. Accordingly, we give the average disturbed times of every node in total 120000 querying times in four topologies when *N*=2000, as shown in Fig.8.

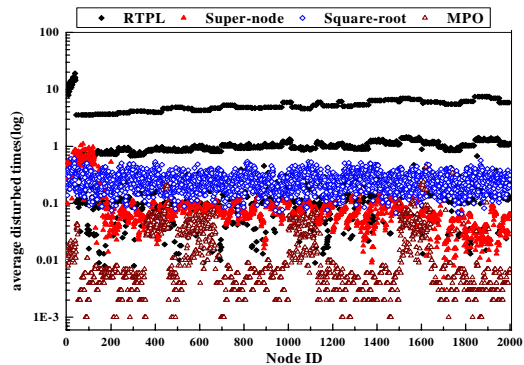

Fig. 8. Disturbed times of every peer

From Fig.8, it is obvious that to each node in MPO, the average disturbed times are less than other 3 topologies, some even equal to 0. Being similar to super-node topology, the average disturbed times of some peers are higher than others in MPO. It is understandable that they are the ICs which are described in detail in [37]. But it is clear that the average disturbed times of the ICs whose loads are the most are farther less than super peers in super-node topology, and near those of the normal node in the latter. It may be also one of the reasons that the cost of MPO is less.

4.2.4 Fault-tolerance and robustness

In this section, we concentrate on discussing the performance of every topology using unrepeated flooding algorithm when there are peers to leave system randomly. Fig.9 shows the performances of topologies, where the *x*-axis is the ratio of the number of randomly leaving nodes to all nodes' number in the system.

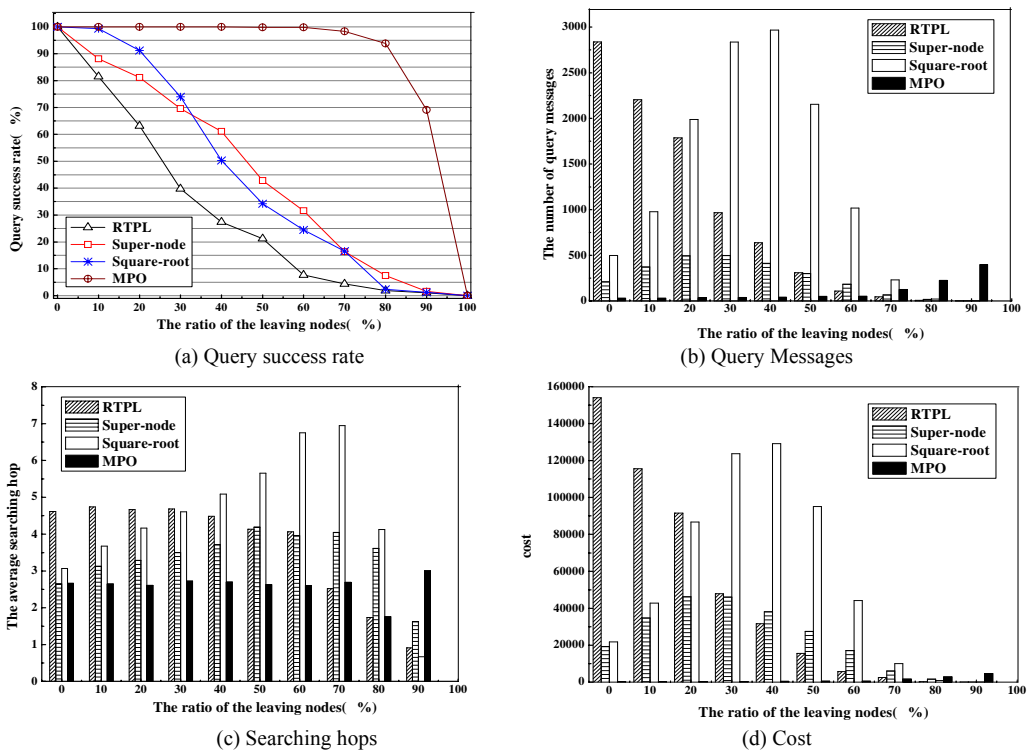

Fig. 9. The different performance when there is randomly leave

It can be seen from the Fig.9 that the results are consistent with the conclusion in theory. Fig.9(a) shows that the query success rate in MPO is very high. Fig.9(c) shows that the average hops in the RTPL, super-node topology and square-root topology all increase when there are nodes randomly leaving. But the average hops in the MPO are almost unchanged and only increase when the condition is very worse (e.g. 90% nodes leaving). In the worse condition, there are too much nodes leaving the system so that the query hops increase when the query is successful. In addition, it is clear in the Fig.9(b), (c) and (d) that each performance in the square-root worsens with the number of leaving nodes increasing. We think the reason is that the formation of square-root topology is based on the query frequency of the files. In square-root topology, the node whose degree is more is consequentially the node whose own files are more popularity. So each aspect of performance in square-root topology is certainly affected when the number of leaving nodes increases. The number of query message and system cost in the MPO when the query is successful don't greatly wave since the nodes leave. It shows that the MPO topology structure has favorable stability.

From Fig.9, it is very different between the intuitive judgement and the real performance of square-root topology. To our intuitive judgement, the query success rate of square-root topology should be far higher than the two latter, while the number of query messages, the query hops and cost should be less than those of the two latter. However, from Fig.9, we find it is so different from our intuition, e.g. the query success rate of square-root topology is similar to the two latter. This is decided by the characteristic of the square-root topology forming. The physical topology of the square-root is robust enough because of its high average connected degree. But the files on the nodes whose degree are high have higher popularity. When such nodes leave the system more and more, the query success rate will decrease. It reflects that the square-root topology doesn't consider the matching between the physical topology and the logical structure. It is clear through the detailed analyzing that the MPO is better than other three topologies in both the physical topology and logical structure.

In simulations (not shown in the paper), we also have evaluated the affects of mechanism such as incentive mechanism for preventing negative behaviors. Considering factors including query success rate, query hops and query message number and system cost whether using incentive mechanism or not, we find that those mechanisms are effective so that the performance of the system applying the incentive mechanism is better and better than that of the system not applying the incentive mechanism as time increasing. That conclusion is shown by many researches for negative behavior preventing mechanism. However, we also find that the response of the square-root topology to those mechanisms is the most obvious. Because square-root topology builds node's degree according to the query frequency of the node's files. If the popularity of the node's file is high and the file is continually visited, the nodes connect with it are more. So considering the Zipf distribution of the file distribution and query frequency, if the nodes whose own files' visited frequencies are high adopt mostly non-cooperation attitude, the failing rate to query this file is great. This is obvious in the experiment. In addition we say from the simulation results that the query hops to successful find in MPO are still less even if there are large numbers of non-cooperation nodes. Thus, it concludes that a better topology can make the system to increase the tolerance to the non-cooperation action.

## 5. Conclusion

In the paper, we investigated the challenges facing autonomic communication, and combined characteristic of unstructured P2P overlay with the merits of a structured P2P overlay to present a Maya pyramid structure P2P overlay (MPO) for satisfying these challenges. We found this structure could: (1) keep topology, node insertion and update stable, whilst affecting only several nodes in a very small locality of the related space, (2) transfer and process information at low cost, using small world characteristics to improve efficiency and decrease the cost of routing, (3) as well as being robust, a tri-Information Center (Tri-IC) mechanism was presented in order to improve robustness by alleviating the load of cluster heads in a hierarchical P2P overlay, (4) a source ranking mechanism was proposed in order to discourage free riding and whitewashing and to encourage frequent information exchanges between peers. (5) a 3D unique ID structure was presented in the unstructured P2P overlay. This will guarantee anonymity in routing, and will be, not only more efficient because it applies the DHT-like routing algorithm in the unstructured P2P overlay, but also more adaptive to suit AC. Simulation experiments proved that our overlay is robust, highly efficient and of a low-cost.

**Acknowledgements**

This research is supported by Doctoral Program Foundation of Institutions of Higher Education of China under Grant No. 20100162120015, and by the freedom explore Program of Central South University under Grant No. 201012200183.